\documentclass[twocolumn,twoside,slac_two]{revtex4}
\usepackage{graphicx}
\usepackage{fancyhdr}
\begin{document}

\title{Fermi's Mystery Sources: Methods for Classification and Association}

%

\author{Elizabeth C. Ferrara, Roopesh Ojha, Maria Elena Monzani, Nicola Omodei}
\affiliation{NASA/ Goddard Space Flight Center, Code 661, Astroparticle Physics Laboratory
Greenbelt, MD 20771}
\author{on behalf of the \textsl{Fermi} Large Area Telescope Collaboration}

\begin{abstract}
Unassociated Fermi-LAT sources provide a population with discovery potential.  We discuss efforts to find new source associations for this population, and summarize the successes to date. We discuss how the measured gamma-ray properties of associated LAT sources can be used to describe the gamma-ray behavior of more-numerous source classes. Using classification techniques exploiting only these gamma-ray properties, we separate the LAT 2FGL catalog sources into pulsar and AGN candidates.
\end{abstract}

\maketitle

\thispagestyle{fancy}


\begin{figure*}[t]
\centering
\includegraphics[width=135mm]{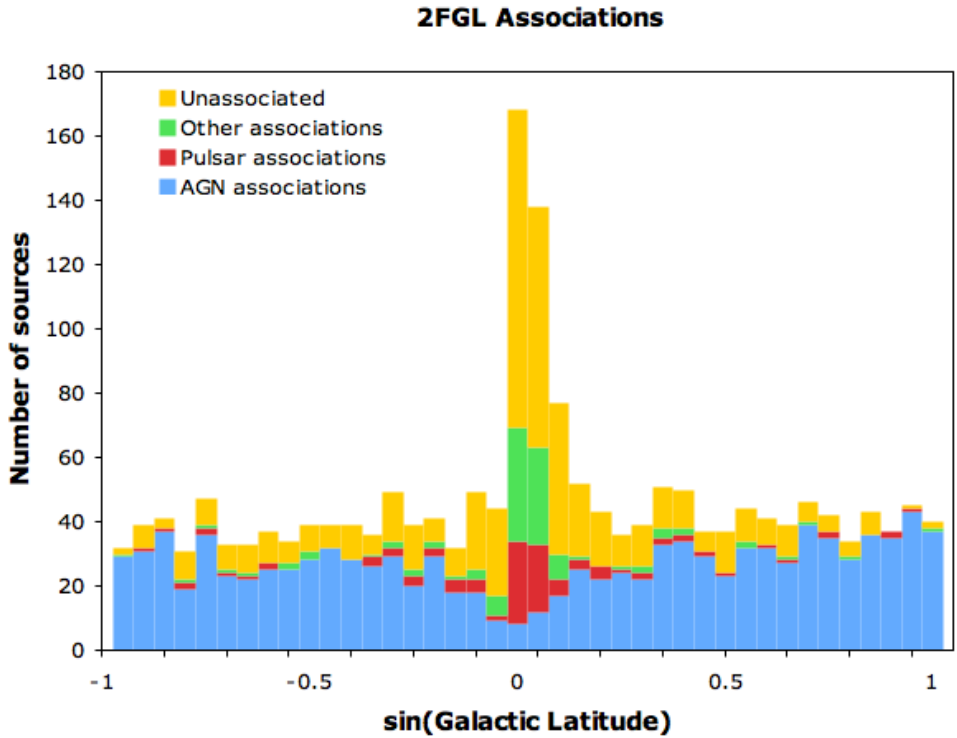}
\caption{2FGL source associations by type as a function of Galactic Latitude. AGN are expected to be isotropic, whilc Galactic sources (pulsars, SNRs, and PWNe) should be concentrated along the Galactic plane. The 2FGL unassociated sources have a large concentration in the Galactic Plane, but are present at all Galactic Latitudes.} \label{2FGL_Assocs_sinb.eps}
\end{figure*}

\section{Introduction}
The gamma-ray sky as seen by the Fermi-LAT after 2 years of observations contains almost 1900 sources. Yet nearly a third of these sources show no clear association with an object belonging to a known gamma-ray emitting class. To investigate the nature of these sources, a number of multi-wavelength initiatives are observing the most promising sources in X-ray, optical, and radio wavebands. In addition, archival searches and correlations with other, less-likely source catalogs provide insights into possibilities for the origin of these sources. We discuss the gamma-ray properties of these mystery sources, the methods being used to investigate them, and summarize the results to date of these various initiatives.

\section{Source Association in the LAT Catalog}
Associating Fermi LAT sources with counterparts of interest is performed using a Bayesian probability based on the position match and the chance coincidence in a given direction. The most likely source classes have been considered~\cite{Abdo2011}:
\begin{itemize}
\item Blazars (BL Lacs, FSRQs, etc.) $\sim 58$\%
\item Other AGN (Seyferts, Radio Galaxies, etc.) $>2$\%
\item Pulsars and binaries (HMXBs, LMXBs, etc.) $\sim 6$\%
\item Other Galactic Sources (SNRs, PWNe, Globular Clusters, etc.) $\sim 4$\%
\end{itemize}
Even after such searches, $\sim 30$\% of Fermi LAT source detections remain unassociated with one of these potential counterparts. These sources represent areas of new discovery.  The result of our classification process is shown in Figure~\ref{2FGL_Assocs_sinb.eps}.

At this time, there is no clear indication of a significantly numerous new class of gamma-ray emitters in the Fermi LAT dataset~\cite{Ackermann2011a}. However, follow-up observations of these sources in other wavebands have provided a number of new discoveries.

In addition to searching for multi-wavelength counterparts, the gamma-ray properties alone have provided clues to the likely source type for a number of unassociated sources.

\subsection{2FGL Intrinsic Parameters}
Intrinsic source parameters~\cite{Abdo2011} for unassociated sources can be compared against patterns of known classes. For 2FGL, the primary intrinsic parameters for separating pulsars from AGN are variability index, significance of spectral curvature, flux and hardness ratios.

\begin{figure*}[t]
\centering
\includegraphics[width=135mm]{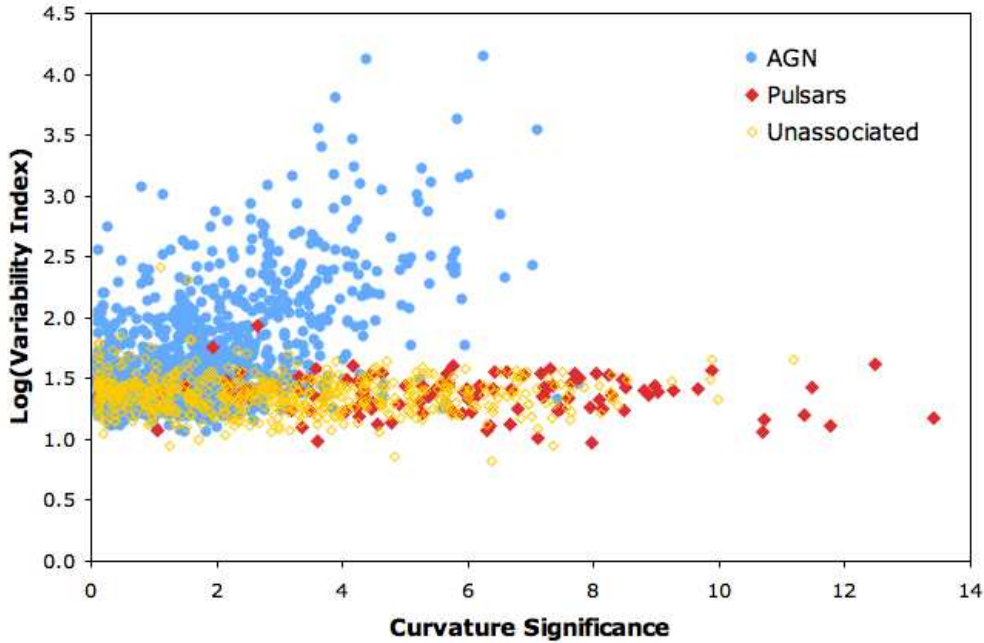}
\caption{Relationship of the Variability Index and Curvature Significance for different object classes.} \label{VI_vs_Curv.eps}
\end{figure*}

Variability in the gamma-rays is a signature of blazars (Figure~\ref{VI_vs_Curv.eps}), with only rare exception. In 2FGL, the variability indicator (TSvar) was provided by a likelihood ratio method which is distributed like $\chi^{2}$ with 23 degrees of freedom (for the 24 months in the data set). 

The 2FGL spectral analysis compared a simple power-law model with a curved log-parabola spectral model. For sources where the change in significance from the use of the curved model is significant, the curved model was used for the global fit. A significant curve to the spectrum is a signature of gamma-ray pulsars. 

The flux for each 2FGL source was determined in five separate energy bands. Hardness ratios and color difference both sample spectral shape without requiring a full fit. However, faint sources are often undetected in multiple bands, making these values somewhat less useful. 

\subsection{Searching for Pulsars in 2FGL}
Currently detected LAT pulsars are typically:
\begin{itemize}
\item Non-variable (TSvar $<41.6$ in 2FGL)
\item Average cutoff at 2.3 GeV
\item Hard F0.1-0.3/F0.3-1 flux ratio ($\sim 1.4$) below the cut-off
\end{itemize}
These parameters can be used to define sources that look ``pulsar-like'' in the LAT and target them for follow-up observations or further study.

Pulsar searches in the LAT data use several different methods:
\begin{enumerate}
\item Previously-known energetic pulsars discovered in radio or X-ray are observed to provide a current ephemeris (a service provided by the Pulsar Timing Consortium). LAT team searchers have used these ephemerides to fold the LAT data and have detected pulsations for 38 pulsars.
\item Blind searches~\cite{Saz-Parkinson2010} use the arrival times of events from previously unassociated LAT sources to look for periodicities. To date, 35 new radio-faint/quiet gamma-ray pulsars have been identified from such sources~\cite{Saz-Parkinson2010,Pletsch2011}.
\item In addition, a group of radio astronomers (the Pulsar Search Consortium) is observing pulsar-like unassociated sources to search for for previously unknown radio pulsars. Where possible, these searchers use the positions of X-ray sources in the LAT error ellipse as their targets.  By following up on numerous unassociated sources, this group has discovered 33 new millisecond and 2 new young pulsars~\cite{Ackermann2011a}.
\end{enumerate}

\subsection{Searching for Blazars}
Typical blazar characteristics in the LAT are:
\begin{enumerate}
\item Time-variable (TSvar $> 41.6$)
\item Power-law or broken power-law spectral shape
\item High probability of association with a known blazar~\cite{Ackermann2011b}
\end{enumerate}
Sources with the first two characteristics can be considered ``blazar-like'', making them prime targets for AGN follow-up. Such follow-up is often important for sources at low-Galactic latitude, as that is a region often avoided by AGN surveys, making AGN under-represented in that region.

New blazars have been found in LAT sources by:
\begin{enumerate}
\item Significant non-periodic variability in an unassociated LAT source is indicative of a probable blazar.
\item Follow-up observations of LAT unassociated sources with the VLA have discovered a number of new flat-spectrum sources. This method uses known catalogs to find radio sources within the LAT error ellipse. Observations in several bands of those sources can determine if one of them is a flat-spectrum source, and thus a likely blazar. 
\end{enumerate}

A new program of radio observations using the Ceduna Hobart Interferometer (CHI), as part of the TANAMI (Tracking Active Galactic Nuclei with Austral Milliarcsecond Interferometry~\cite{Ojha2010}) program, is working to fill in the low-Galactic latitude regions, as well as following up on southern hemisphere unassociated sources unreachable by the VLA (at declinations less than -40¼).  If results are comparable to the VLA follow-up program, we expect to detect 35-40 new blazars using this technique.

\section{Classification Using Intrinsic Properties}
In order to classify unassociated point sources from the 2FGL Fermi-LAT catalog~\cite{Abdo2011} as likely blazars and likely pulsars, we compared their measured gamma-ray parameters to the same parameters for the associated sources. 1077 sources associated with AGN and 108 sources with pulsar associations were used as a training sample for a Classification Tree analysis~\cite{Ackermann2011a}. The intrinsic parameters that provide significant signal to separate the two source classes (in order of decreasing significance) are:

\begin{itemize}
\item Variability TS and Spectral Index
\item Curvature Significance and Color difference
\item High-energy flux (3-10 and 10-100 GeV bands)
\item Low-to-High energy hardness ratio
\end{itemize}

\begin{figure*}[t]
\centering
\includegraphics[width=135mm]{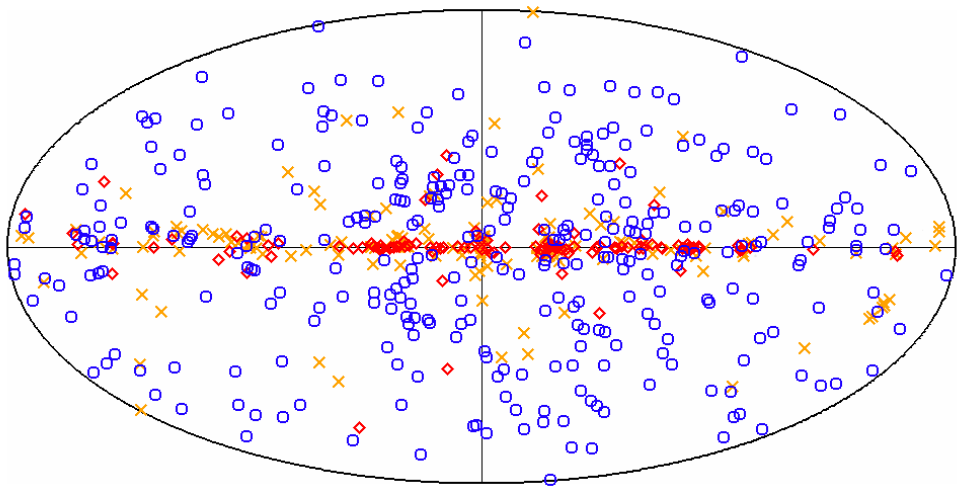}
\caption{Classification Tree Results: Spatial distribution of 2FGL unassociated sources in Galactic coordinates.} \label{CT_results_spatial.eps}
\end{figure*}

\begin{figure*}[t]
\centering
\includegraphics[width=135mm]{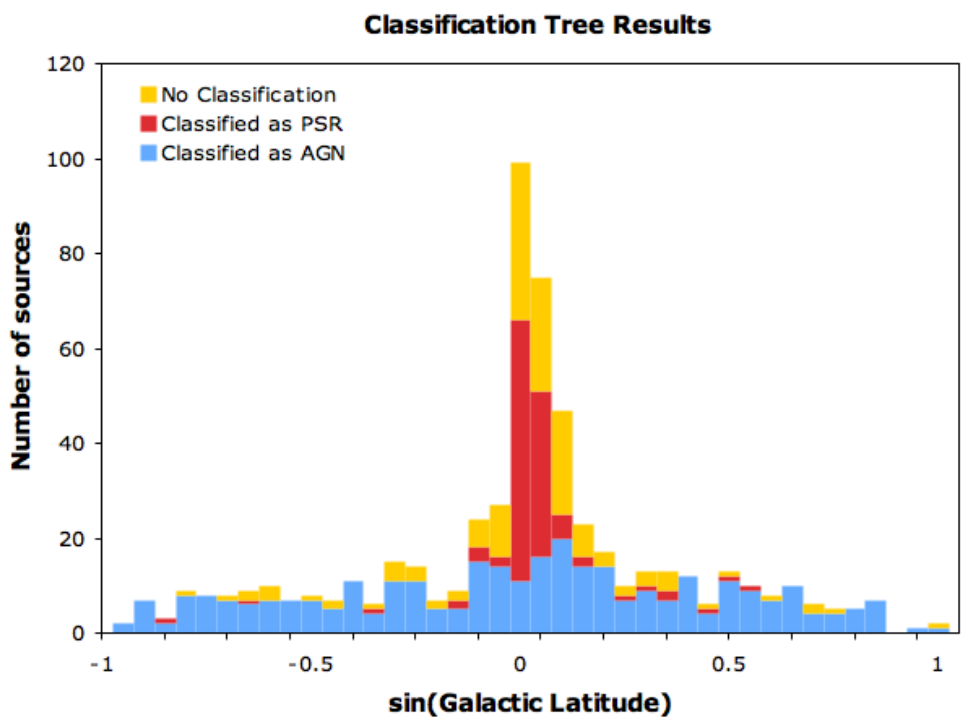}
\caption{Classification Tree Results of 2FGL unassociated sources as a function of Galactic latitude. Pulsar candidates are clustered in the Galactic plane, and AGN candidates are more isotropically distributed.} \label{CT_results_sinb.eps}
\end{figure*}

We then used the model derived from the Classification Tree analysis to classify the 2FGL unassociated source population (Figure~\ref{CT_results_spatial.eps}) into 315 AGN candidates and 114 pulsar candidates. 144 sources were unable to be classified by this method (Figure~\ref{CT_results_sinb.eps}).

The results from this classification technique have been used to help inform the next set of multi-wavelength observations searching for new pulsars, blazars, and other gamma-ray emitting objects. And the results from those searches will help refine the technique for future catalogs. 

There remains the group of 144 sources that could not be easily classified as either a likely pulsar or a likely AGN. For the most part, these are non-varying sources that are too faint in gamma-rays to have a well-determined spectral shape. One possibility is that these are faint AGN that have not displayed an outburst during the two years of Fermi data included in the 2FGL catalog. 

\bigskip 
\begin{acknowledgments}
This research was funded in part by NASA through Fermi Guest Investigator grants NNH09ZDA001N and NNH10ZDA001N. This research was supported by an appointment to the NASA Postdoctoral Program at the Goddard Space Flight Center, administered by Oak Ridge Associated Universities through a contract with NASA.
\end{acknowledgments}

\bigskip 

\begin{thebibliography}{9}   

\bibitem{Abdo2011} Abdo, A. A. et al., "Fermi Large Area Telescope Second 
Source Catalog", 2011, eprint arXiv:1108.1435
\bibitem{Ackermann2011a} Ackermann, M. et al., "A Statistical Approach to Recognizing Source Classes for Unassociated Sources in the First Fermi-LAT Catalog", 2011, eprint arXiv:1108.1202
\bibitem{Ackermann2011b} Ackermann, M. et al., "The Second Catalog of Active Galactic Nuclei Detected by the Fermi Large Area Telescope", 2011, ApJ, 743, 171 eprint arXiv:1108.1420
\bibitem{Saz-Parkinson2010}Saz-Parkinson, P. M. et al., "Eight $\gamma$-ray Pulsars Discovered in Blind Frequency Searches of Fermi LAT Data", 2010, ApJ, 725, 571
\bibitem{Pletsch2011} Pletsch, H. J. et al., "Discovery of Nine Gamma-Ray Pulsars in Fermi Large Area Telescope Data Using a New Blind Search Method", 2012, ApJ, 744, 105
\bibitem{Ojha2010} Ojha, R. et al., "TANAMI: tracking active galactic nuclei with 
austral milliarcsecond interferometry. I. First-epoch 8.4 GHz images", 2010, A\&A, 
519, A45
\end{thebibliography}

\end{document}